\documentclass[10pt]{article}

\usepackage[english]{babel}
\usepackage{amsmath,amsthm}
\usepackage{amsfonts}
\usepackage{graphicx}
\usepackage{amssymb}
\usepackage[a4paper]{geometry}
\numberwithin{equation}{section}

\begin{document}

\title{Modelling the spreading rate of controlled communicable epidemics through an
entropy-based thermodynamic model}
\author{W.B. Wang, Z.N. Wu\thanks{Corresponding Author, ziniuwu@tsinghua.edu.cn}, Z.M. Cao\\
\small Department of Engineering Mechanics, Tsinghua University, China
\and  R.F. Hu\\
\small School of Electronical and Mechanical Engineering, Xidian University, China}
\date{April 19, 2013}
\maketitle

\begin{abstract}
{\small A model based on a thermodynamic approach is proposed for predicting
the dynamics of communicable epidemics in a city,  when the epidemic is governed by
controlling efforts of multiple scales so that an entropy is associated with
the system. All the epidemic details are factored into a single parameter
that is determined by maximizing the rate of entropy production. Despite the
simplicity of the final model, it predicts the number of hospitalized cases
with a reasonable accuracy, using the data of SARS of the year 2003,  once
the inflexion point characterizing the effect of multiple controlling
efforts is known. This model is supposed to be of potential usefulness since
epidemics such as avian influenza like H7H9 in China this year have the risk to become communicable among human beings.}
\end{abstract}


\section{Introduction}
Starting from November 2002 till the end of May 2003, the severe acute respiratory syndrome (SARS)
had spread widely over the world. Up to the end of May 2003 probable cases have
been reported in 35 countries or regions, and the cumulative number of cases
has reached 8202 by May 26, 2003 according to the report by the World Health
Organization (WHO). SARS in the year 2003 and avian influenza like H7N9 this year received or receive intensive attentions
from all over the world due to its high case-fatality rate. People were
particularly interested in finding the period of the time between infection
and the onset of infectiousness, length of period that patients remain
infectious, further infections that each patient produce, total number of
infections during the epidemic, etc. A large number of publications have
been reported for SARS, many of which have been included in reference books \cite{ref0,ref0-1} and reviews \cite{ref0-2,ref0-3}. Important achievements have been made for the transmission dynamics
using various mathematical models \cite{ref0-a}-\cite{ref0-m} and
reported data from Hong Kong or Canada. Donnelly et al\cite{ref0-a}, Riley
et al \cite{ref0-b} and Lipsitch et al \cite{ref0-c} make use of the
available data for SARS on latent, incubation and infectious periods and
have successfully fitted their models to data describing
the number of cases observed over time. \ The important conclusion is that
if the SARS is uncontrolled, then a majority of people would be infected.
The potential effectiveness of different control measures has been studied
in these references.

Though SARS did not appear again since 2003, there may be other epidemic, such as H7N9 avian influenza occurring actually in China possibly,  spreading in a similar way. Hence the study of
various models for the prediction of SARS and other epidemic  once it occurs is always important.

As assessed by Dye and Gay \cite{ref0-d}, the current mathematical models
are complex, the data are poor, and some big questions such as accuracy of
case reports and heterogeneity in transmission remain. Dye and Gay
anticipated that the next generation of SARS models would have to become
more complex.

It is now evident that SARS and maybe avian influenza in a city can be controlled through multiscale
measures such as medical interventions, public-service announcements,
isolation of people having contact with infected, \ restriction of
individual and social activities, etc. When the interventions to control
a communicable epidemic are intensive and of multiple scales, it would be very diff$^{{\small %
\cdot }}$icult to find all those details of the epidemic
needed by a more complex model. It is thus desired that, under intensive and
multiscale interventions, the global behavior of SARS or avian influenza spread, governed by a
complex and multiscale system, could be roughly predicted without knowing
the epidemic details.

The dynamics of an epidemics is an important topic in biology, medicine,
mathematics and physics and is usually modelled through differential
equations \cite{refm1}-\cite{bcc}, among which is the famous SIR
(susceptible-infected-removed) model. The study on this topic is still \
very active (\cite{ref0-e}-\cite{ref0-m},\cite{ref3}-\cite{ref6},).

Most of the models for epidemics spread rely on differential equations for
the susceptible, infected and removed numbers. Different spread mechanisms
are embedded into the various terms in the differential equations.

In this paper we are interested in the number of hospitalized cases
(cumulative number of cases minus the number of deaths and the number
recovered) and attempt to consider a new approach to predict this number. In
our approach all the mechanisms controlling the spread are factored into a
single parameter. Assuming the system controlling the spread of SARS or similar epidemic is a
thermodynamic one, we define an entropy and determine
the only parameter by using the principle of extreme rate of entropy
production. This allows us to relate the dynamics of the spread to the
information at the inflexion point of the curve
describing the time variation of the number of hospitalized cases. The inflexion point is the date at which the multiple controlling
measures take effect.

The model presented in this paper is based on a simple differential equation
with the spread rate forced to satisfy four constraints (section 2.1). The
model is closed by the use of maximum or minimal rate of entropy production
as the system for spread is assumed to be a thermodynamical one (section
2.2). There is a critical point (date) at which the spread rate turns to
decrease due to the overall role of interventions. The maximum number of
infected individuals and the time at which this maximum occurs can be
related to the number and time corresponding to the critical date (section
2.3). This model is validated against the SARS data of the year 2003  (section 3) for which we are able to follow the history of the
spread.

\section{ Model development}

\subsection{Basic model}

Let $f(t)$ be the number of hospitalized cases, defined
as the cumulative number subtracted from the cumulative number of deaths and
recovery ones since death and recovery are also parts of actions in the
thermodynamical system. Then the rate of increase (decrease) is proportional
to the number at the previous day,

\begin{equation}
\frac{df(t)}{dt}=\alpha (t)f(t)  \label{001}
\end{equation}%
with the roles of all the controlling mechanisms factored into the parameter
$\alpha (t)$. Knowing the exact expression of $\alpha (t)$ requires the
knowledge of all the details of the epidemic and the coupling with other
differential equations. The essential idea in our method to find $\alpha (t)$ \ is to ignore any details but to use a
thermodynamic approach. It is to be remarked that this parameter must be
subjected to the following four constraints:

1) The parameter $\alpha (t)$ must have the dimension of $t^{-1}$, i.e. $%
\alpha (t)\sim t^{-1}.$

2) At the initial stage there is an exponential increase for regular spread
to start (since at the initial stage the number is near zero), i.e., $\alpha
(0)\rightarrow \infty $.

3) With the strong and active interventions the rate must decrease at a
given day $t=L$ which will be called the \emph{inflexion point}
(date). Mathematically this amounts to say that $\frac{d^{2}f(t)}{dt^{2}}$
vanishes at $t=L$, i.e.,

\[
\frac{d\alpha }{dt}+\alpha ^{2}=0,t=L
\]

4) There must be a maximum for $f(t)$, say at ${\normalsize t=D}$, for which
we have $\alpha (t)=0$.

We assume that the virus causing an epidemic is constantly active (high temperature
or intrinsic lifetime constraint would make the epidemic disappear suddenly, but
this is not considered here) so that $\alpha (t)$ is assumed to be an
analytical function. Also nature would select laws as simple as possible.
The only analytical function that meets the four constraints and that is
simple enough is found to be given by

\begin{equation}
\alpha (t)=\frac{-c\ln (t/D)}{t}  \label{002}
\end{equation}%
where $c$ is a parameter. Inserting (\ref{002}) into (\ref{001}) leads to
the following solution which is just the log-normal function,

\begin{equation}
\bigskip f(t)=\frac{k}{\sqrt{2\pi }\sigma t}\exp \left( -\frac{(\ln t-\mu
)^{2}}{2\sigma ^{2}}\right)   \label{eq1}
\end{equation}%
Here $k$ is a proportion constant, ${\normalsize \mu =\ln D+\sigma ^{2}}$,
and ${\normalsize \sigma }$ is to be determined in the following through the
use of the principle of extreme rate of entropy production.

\subsection{Principle of extreme rate of entropy production}

The principle of extreme rate of entropy production can be found in \cite%
{ref8}. This has been successfully \ used to obtain the distribution of
droplet production during its impingement on solid walls \cite{ref8bis}.
Certainly, the width of the curve $f(t)\propto t$ can be characterized by $%
{\normalsize \sigma }$. The wider the curve is, the larger is the (Shannon)
entropy. \ The intrinsic spread mechanism of virus and the large mixing
activity of the population tend to make the curve wider (so ${\normalsize %
\sigma }$ larger). However, the medical and social interventions to control
the epidemic constitute a dissipation mechanism which would prohibit the curve to
become infinitely wide (${\normalsize \sigma }$ infinitely large). The width would cease to increase when the
maximum dissipation rate is reached. Maximum dissipation rate corresponds to
extreme rate of entropy production, which again corresponds to
\begin{equation}
\frac{d^{2}S({\normalsize \sigma ,}\eta )}{d{\normalsize \sigma }^{2}}=0
\label{eq23}
\end{equation}%
Here $S({\normalsize \sigma ,}\eta )$ is the Shannon entropy def$^{{\small %
\cdot }}$ined as
\[
S(\sigma ,\eta )=-\int_{0}^{\infty }F(t)\ln F(t)dt^{\eta }
\]%
where $F(t)=t^{1-\eta }f(t)$ with $\eta =3$ in the usual entropy def$^{%
{\small \cdot }}$inition. Integration leads to
\[
S(\sigma ,\eta )=\eta \left( \ln \left( \sqrt{2\pi }\sigma \right) +\eta
\left( {\normalsize \ln D+\sigma ^{2}}\right) +\frac{1}{2}\right)
\]%
so that (\ref{eq23}) holds if and only if

\begin{equation}
\sigma =\frac{1}{\sqrt{2\eta }}\approx
\begin{array}{cc}
0.408 & \text{ for }\eta =3%
\end{array}
\label{eq3}
\end{equation}

\subsection{Maximum number of hospitalized cases}

Inserting (\ref{eq1}) into $\left. \frac{d^{2}f(t)}{dt^{2}}\right\vert
_{t=L}=0$ yields the following relationship between $D$ and $L$

\[
D=L\exp \left( \frac{1}{2}\sigma ^{2}+\frac{1}{2}\sqrt{4\sigma ^{2}+\sigma
^{4}}\right)
\]%
Using (\ref{eq1}), $f(D)$ is related to $f(L)$ by
\[
f(D)=f(L)\exp \left( -\sigma ^{2}-\frac{1}{2}\sqrt{4\sigma ^{2}+\sigma ^{4}}+%
\frac{1}{2}\left( \frac{3}{2}\sigma +\frac{1}{2}\sqrt{4+\sigma ^{2}}\right)
^{2}\right)
\]%
With $\sigma $ given by (\ref{eq3}), we have

\begin{equation}
\left. \frac{D}{L}\right\vert _{\eta =3}=1.\,\allowbreak 649,\left. \frac{%
f(D)}{f(L)}\right\vert _{\eta =3}=\allowbreak 2.\,\allowbreak 120
\label{eq6}
\end{equation}

\subsection{Initial date for regular spreading}

Once we know the inflexion date, it is crucial to determine when
is the initial date for regular spreading of the epidemic. In other words, we must
know the number $L$ (cumulated days to reach the inflexion point counting from the initial date). This can be done by using the
rate of increase $\frac{df(t)}{dt}$ at $t=L$. A simple calculation using (%
\ref{eq1}) yields

\[
\left. \frac{df(t)}{dt}\right\vert _{t=L}=-\frac{1}{\sigma ^{2}L}f(L)\ln
\frac{L}{D}
\]%
which yields
\begin{equation}
L=\left( \frac{1}{2}+\frac{1}{2}\sqrt{\frac{4}{\sigma ^{2}}+1}\right) \frac{%
f(L)}{\left. \frac{df(t)}{dt}\right\vert _{t=L}}=\frac{3f(L)}{\left. \frac{%
df(t)}{dt}\right\vert _{t=L}}  \label{eqll}
\end{equation}

\section{Application and validation of the model}

\subsection{Use of the model}

The model is used as follows.

Step 1 (data recording). Using the reported data we determine the number $F$
at the inflexion date (the date that $\frac{df(t)}{dt}$
tends to decrease). Also determine $\left. \frac{df(t)}{dt}\right\vert
_{t=L} $ by using the reported date. Determine the proportion constant $k$
in (\ref{eq1}) by setting $f_{L}=f(L)$. Then use (\ref{eqll}) to determine $%
L $.

Step 2 (Prediction). Once $L$ and $f(L)$ are known, use (\ref{eq6}) to
predict $D$ and $f(D)$ and plot the curve $f(t)\sim t$ using eq (\ref{eq1})
to predict the number $f(t)$ for $L<t$.

Hence it is essential to determine the inflexion point.
Specifically, this is done as follows. We record the
reported number $f(t)$ for each day and draw the curve $g(t)=g(t)-g(t-1)$.
Once we observe that $g(t)$ reaches a peak (denoted as $G$) at $t=L$, then $%
L $ is considered as the inflexion point. However,
special cautions must be made.

(a) in the early period of the epidemic, it is possible to have report delay
of cases so that a false peak would occur.

(b) for a city or region where the cumulative number of cases remains always
small, it is diff$^{{\small \cdot }}$icult to observe a clear peak. In this
case this approach does not apply.

(c) There is also a possibility to have multiple inflexion points due to new outbreaks, as is in the case of Hong Kong,
Singapore and Canada.

Numerically, $L$ is calculated as
\begin{equation}
L=3\overline{F}/G  \label{eqlll}
\end{equation}%
where $\overline{F}=(2F-G)/2$ is the number of $f$ averaged over two
consecutive dates (at and before the inflexion date).
Still using the log-normal function, we can relate the maximum $H=f(D)$ and
the date $D$ to $F$ and $G$ by\ $H\approx 2.12\overline{F}$ and $D\approx L+2%
\overline{F}/G$ where none of the constants depends on the details of the
epidemic.

\begin{figure}
\begin{minipage}{0.45\textwidth}
\includegraphics[width=\textwidth]{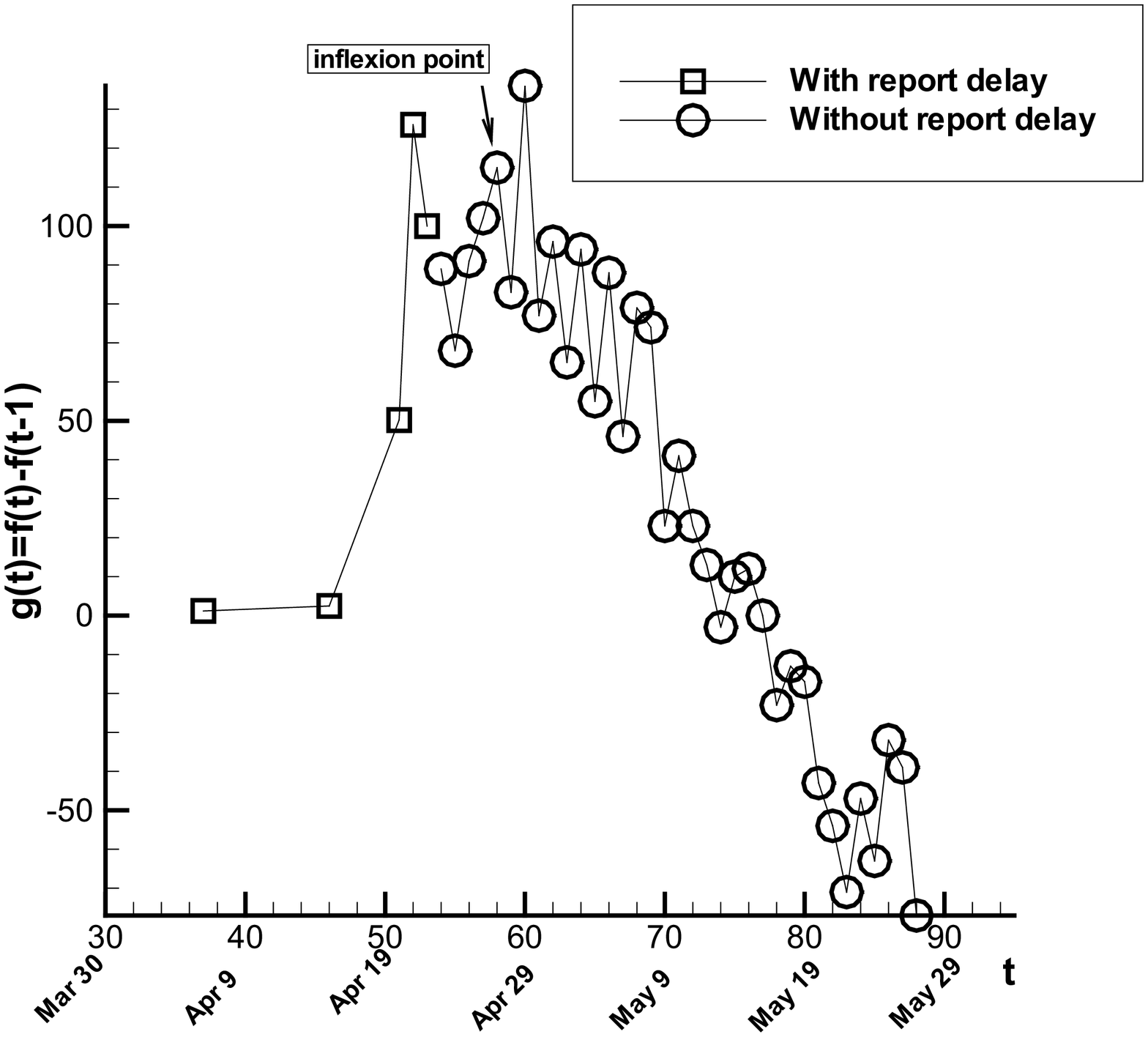}%
\caption{Inflexion point for Beijing. Note that the first peak is
not an inflexion point but is simply due to report delay in the early period.}
\label{fg1}
\end{minipage}
\hfill
\begin{minipage}{0.45\textwidth}
\includegraphics[width=\textwidth]{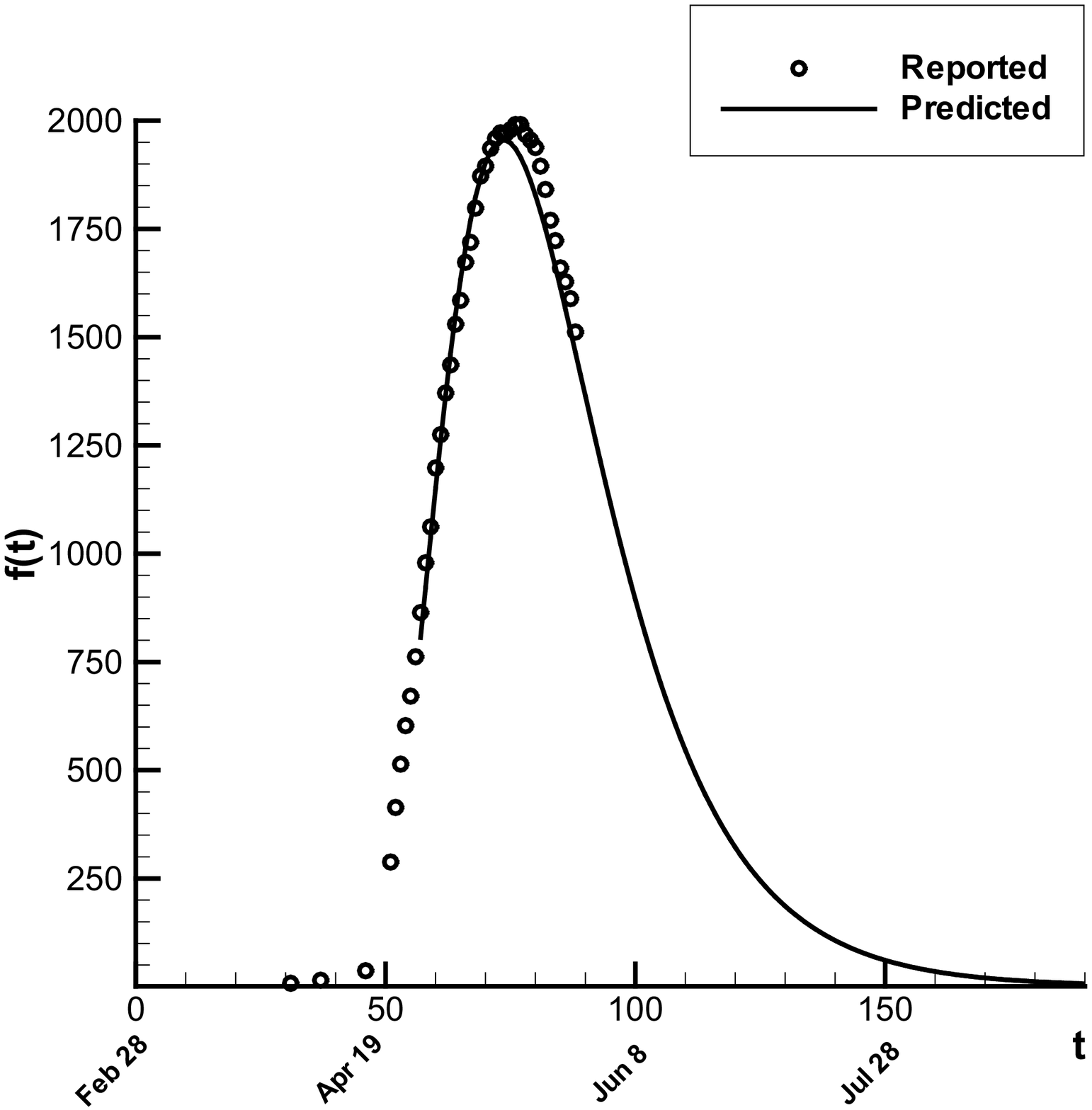}%
\caption{Time history of the number of hospitalized cases for Beijing.}
\end{minipage}
\end{figure}

%
%

\subsection{Test of the model for SARS in 2003}

First consider Beijing. Using the reported date as shown in Fig \ref{fg1}, we identify April 27 to be the critical date since $%
\frac{df(t)}{dt}$ experiences an evident decrease after that date (we also
observe a decrease before April 25, but that decrease is due to the report
delay). Using the reported date we have $f(L)=980$ and $\left. \frac{df(t)}{%
dt}\right\vert _{t=L}=116$. Hence \ $L=24$ according to (\ref{eqlll}). This
means that the initial date for irregular spread is April 3. Using \ (\ref%
{eq6}) we predict $D$ and $f(D)$ to be $D=42$ (May 13) and $f(D)=1955$,
while according to the report, $D=44$ (May 15) and $f(D)=1991$. The
predicted curve $f=f(t)$ follows well the curve, as can be seen in Fig. \ref{fg2}.

\begin{figure}
\begin{minipage}{0.45\textwidth}
\includegraphics[width=\textwidth]{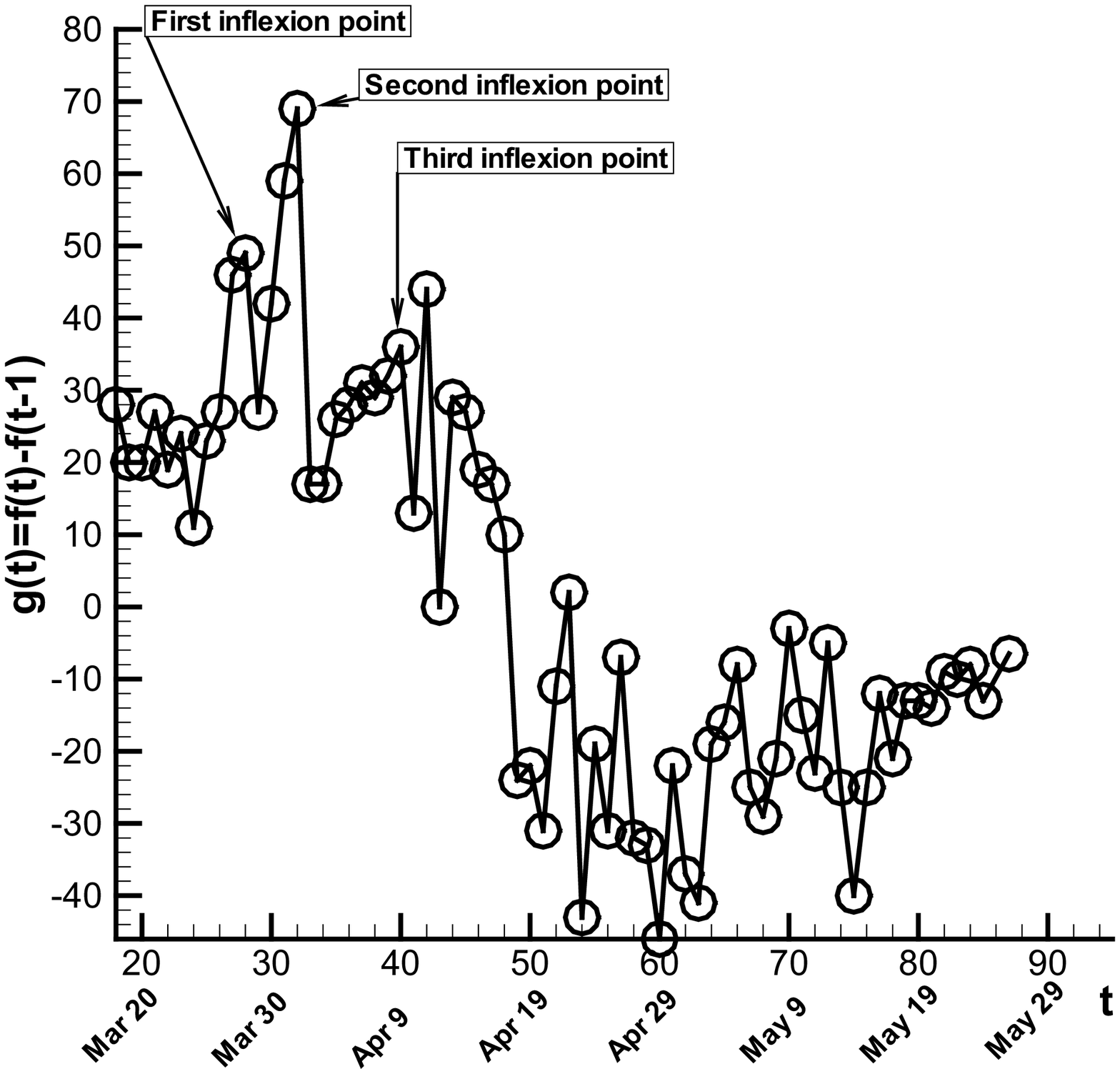}%
\caption{For Hong Kong we observe three distinct inflexion points.}
\label{fg3}
\end{minipage}
\hfill
\begin{minipage}{0.45\textwidth}
\includegraphics[width=\textwidth]{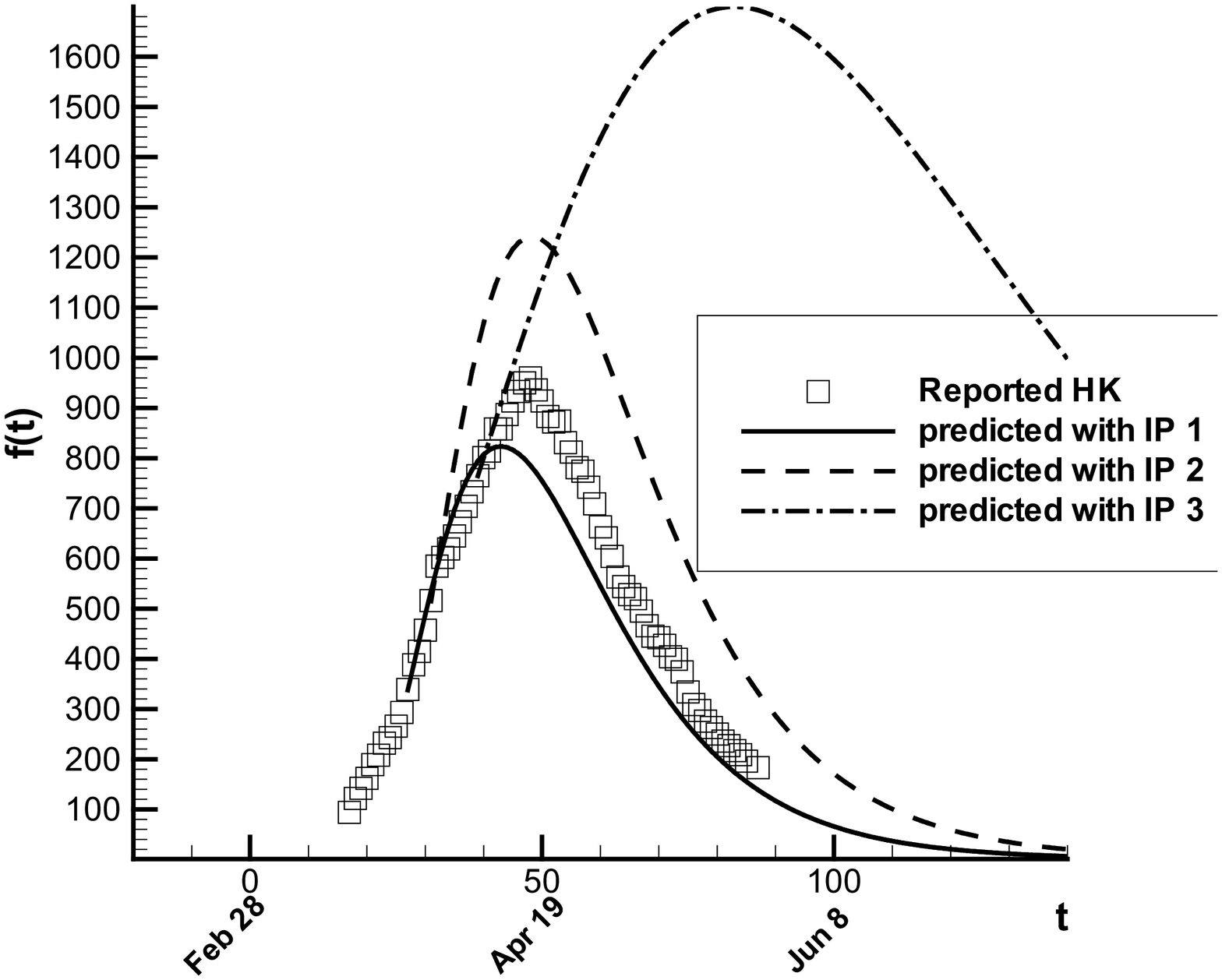}%
\caption{The prediction using the information at the three inflexion points (IP1,
IP2, IP3) shows that the predicted curve using the first inflexion point is the closest to the
reported data.}
\label{fg4}
\end{minipage}
\end{figure}

%

For Hong Kong, we observe three distinct inflexion
points as can be seen in Fig \ref{fg3}. The prediction
using the information at the three inflexion points (IP1, IP2, IP3) show that the predicted curve using the first inflexion point is the closest to the reported data
(Fig. \ref{fg4}). More details can be seen in Table 1.
In Table 1, when there are multiple inflexion points, as
is in the case for Hong Kong, Singapore and Canada, we use the information
at the first inflexion point. In the case of Singapore
and Canada, there are two maximums but we give information only for the first one.

\begin{table}
\begin{center}
\caption{Predicted number $H$ and date $D$ compared with the reported ones for several cities or regions (Mainland refers to Mainland China).}
\label{tab1}
\begin{tabular}{|l|l|l|l|l|l|l|l|l|}
\hline
{\small Regions} & \multicolumn{3}{|l|}{{\small Inf$^{\cdot }$lexion date}}
& \multicolumn{2}{|l|}{{\small Date }$D$} & \multicolumn{2}{|l|}{{\small %
Maximum }$H$} & {\small Error} \\ \hline
& {\small Date} & $F$ & $G$ & {\small Pred} & {\small Rep} & {\small Pred} &
{\small Rep} &  \\ \hline
{\small Beijing} & {\small Apr27} & ${\small 980}$ & ${\small 116}$ &
{\small May13} & {\small May15} & ${\small 1955}$ & ${\small 1991}$ &
{\small 2\%} \\ \hline
{\small HK} & {\small Mar28} & ${\small 960}$ & ${\small 49}$ & {\small Apr12%
} & {\small Apr14} & ${\small 823}$ & ${\small 960}$ & {\small 14\%} \\
\hline
{\small World} & {\small Apr23} & ${\small 2005}$ & ${\small 222}$ & {\small %
May9} & {\small May12} & ${\small 4015}$ & ${\small 3700}$ & {\small 8.5\%}
\\ \hline
{\small Mainland} & {\small Apr27} & ${\small 1572}$ & ${\small 177}$ &
{\small May14} & {\small May12} & ${\small 3332}$ & ${\small 3068}$ &
{\small 8.6\%} \\ \hline
{\small Hebei} & {\small May4} & ${\small 92}$ & ${\small 17}$ & {\small %
May15} & {\small May13} & ${\small 177}$ & ${\small 161}$ & {\small 8\%} \\
\hline
{\small Singap.} & {\small Mar18} & ${\small 29}$ & ${\small 6}$ & {\small %
Mar27} & {\small Mar23} & ${\small 61}$ & ${\small 60}$ & {\small 1\%} \\
\hline
{\small Canada} & {\small Mar27} & ${\small 49}$ & ${\small 9}$ & {\small %
Apr10} & {\small Apr4} & ${\small 104}$ & ${\small 67}$ & {\small 55\%} \\
\hline
\end{tabular}
\end{center}
\end{table}

\begin{figure}
\begin{minipage}{0.45\textwidth}
\includegraphics[width=\textwidth]{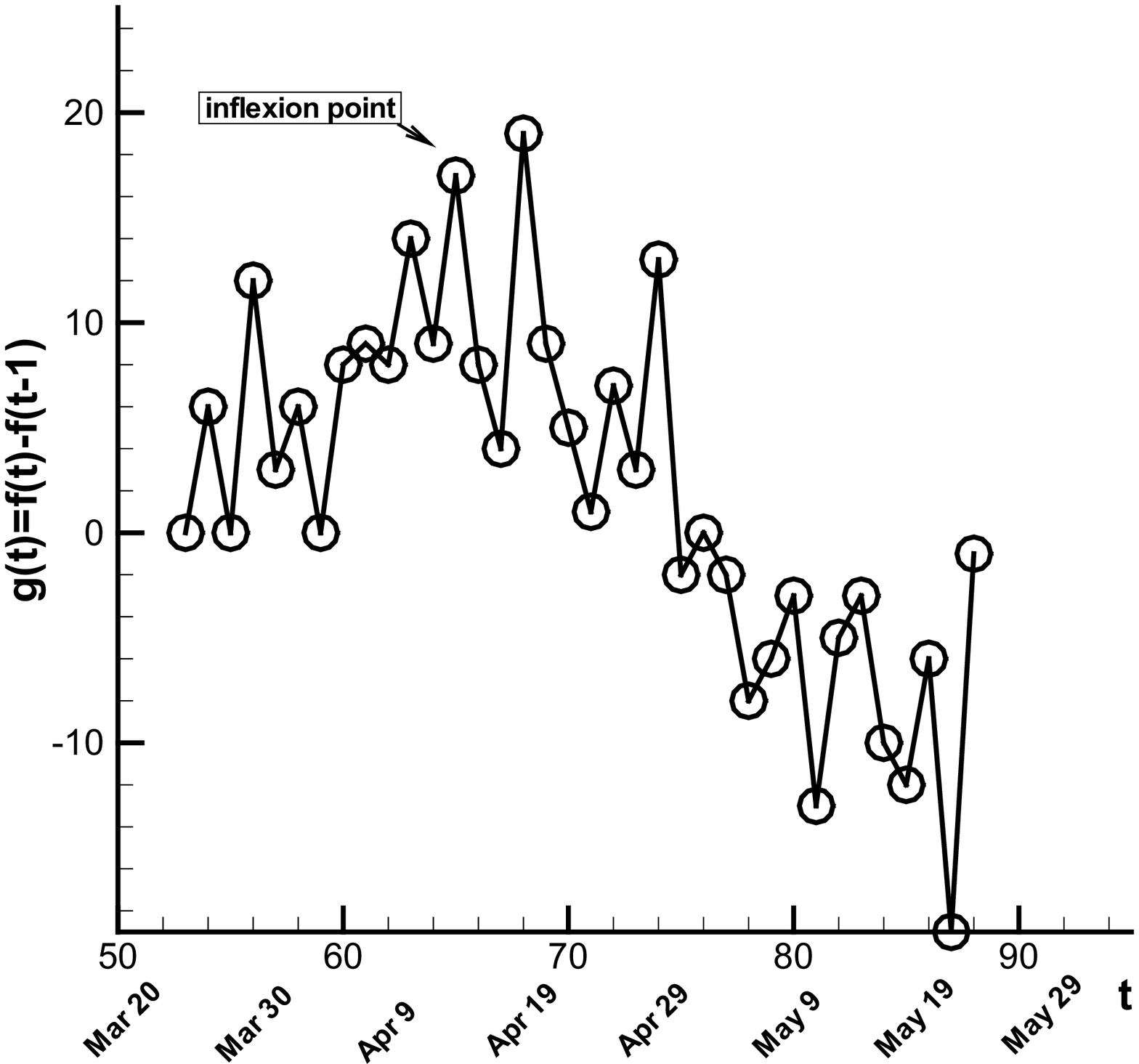}%
\caption{For the Province of Hebei of China the inflexion point is still identifiable.}
\label{fg5}
\end{minipage}
\hfill
\begin{minipage}{0.45\textwidth}
\includegraphics[width=\textwidth]{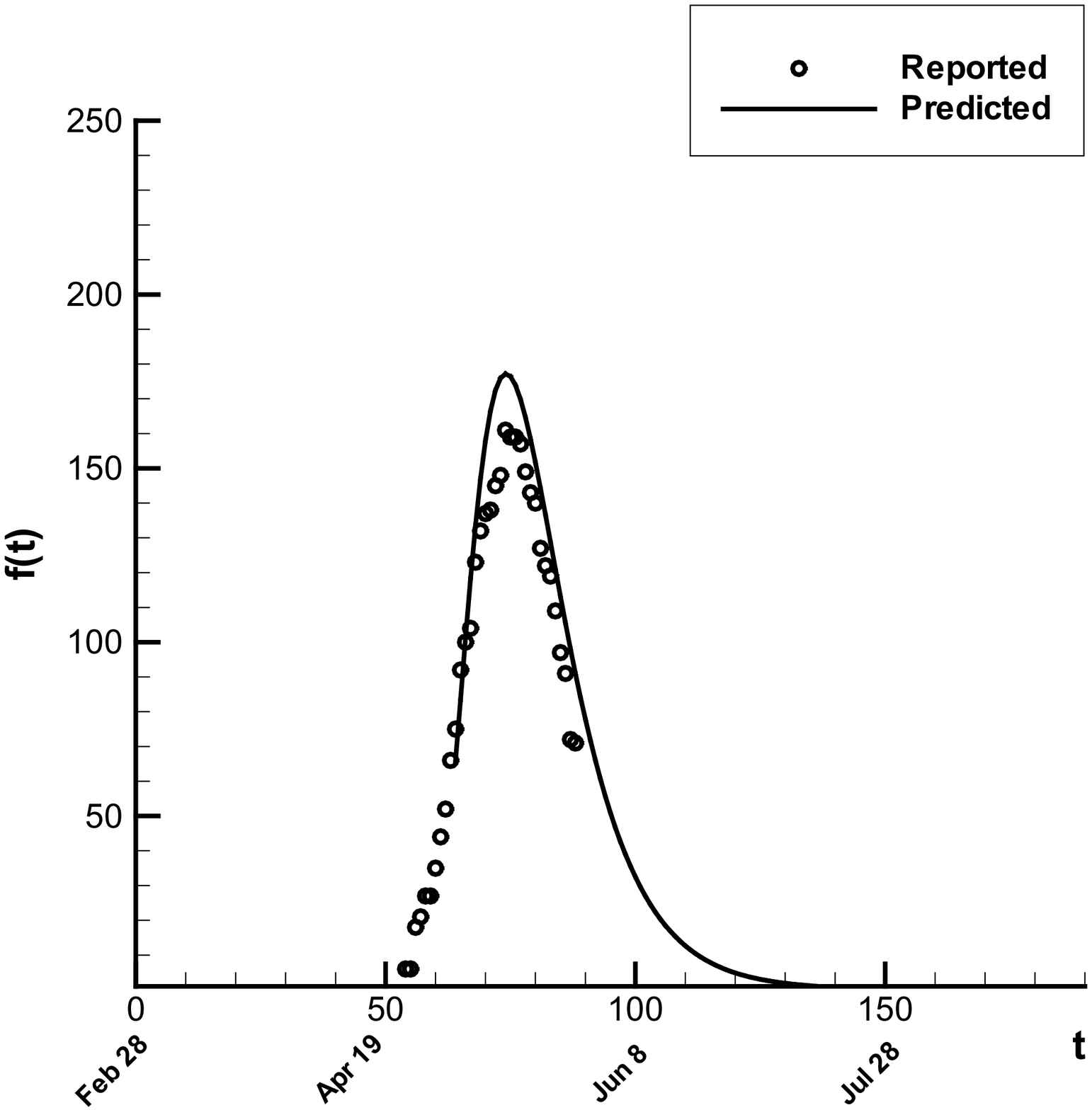}%
\caption{The predicted curve is reasonable as compared to the reported one, though the number of cases in
Hebei is not large.}
\label{fg6}
\end{minipage}
\end{figure}

%

For Hebei, the number of cases is not large. But the prediction still works
very well (Fig. \ref{fg5}, Fig. \ref{fg6}).

\begin{figure}
\begin{minipage}{0.45\textwidth}
\includegraphics[width=\textwidth]{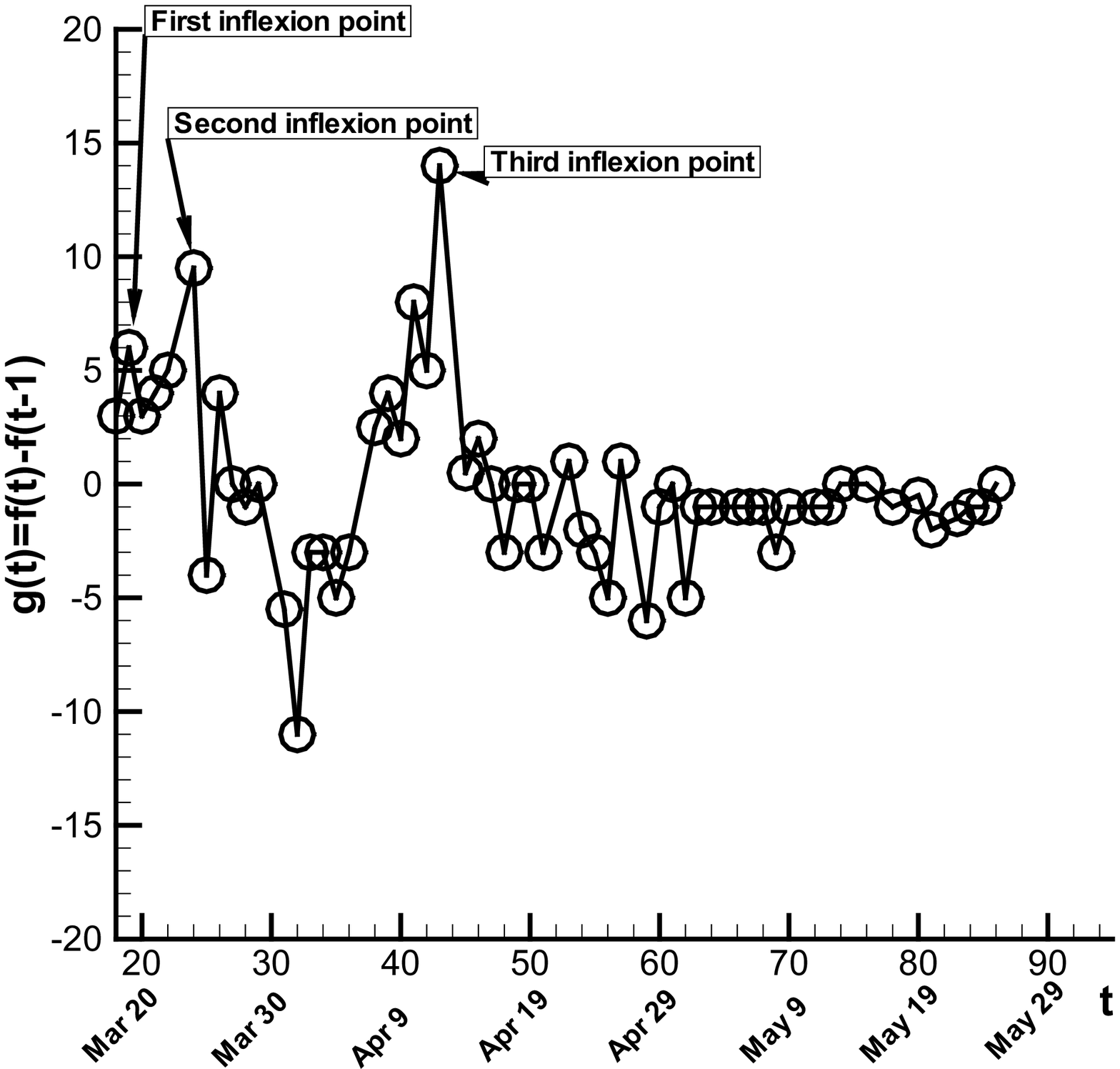}%
\caption{Inflexion points for Singapore.}
\label{fg7}
\end{minipage}
\hfill
\begin{minipage}{0.45\textwidth}
\includegraphics[width=\textwidth]{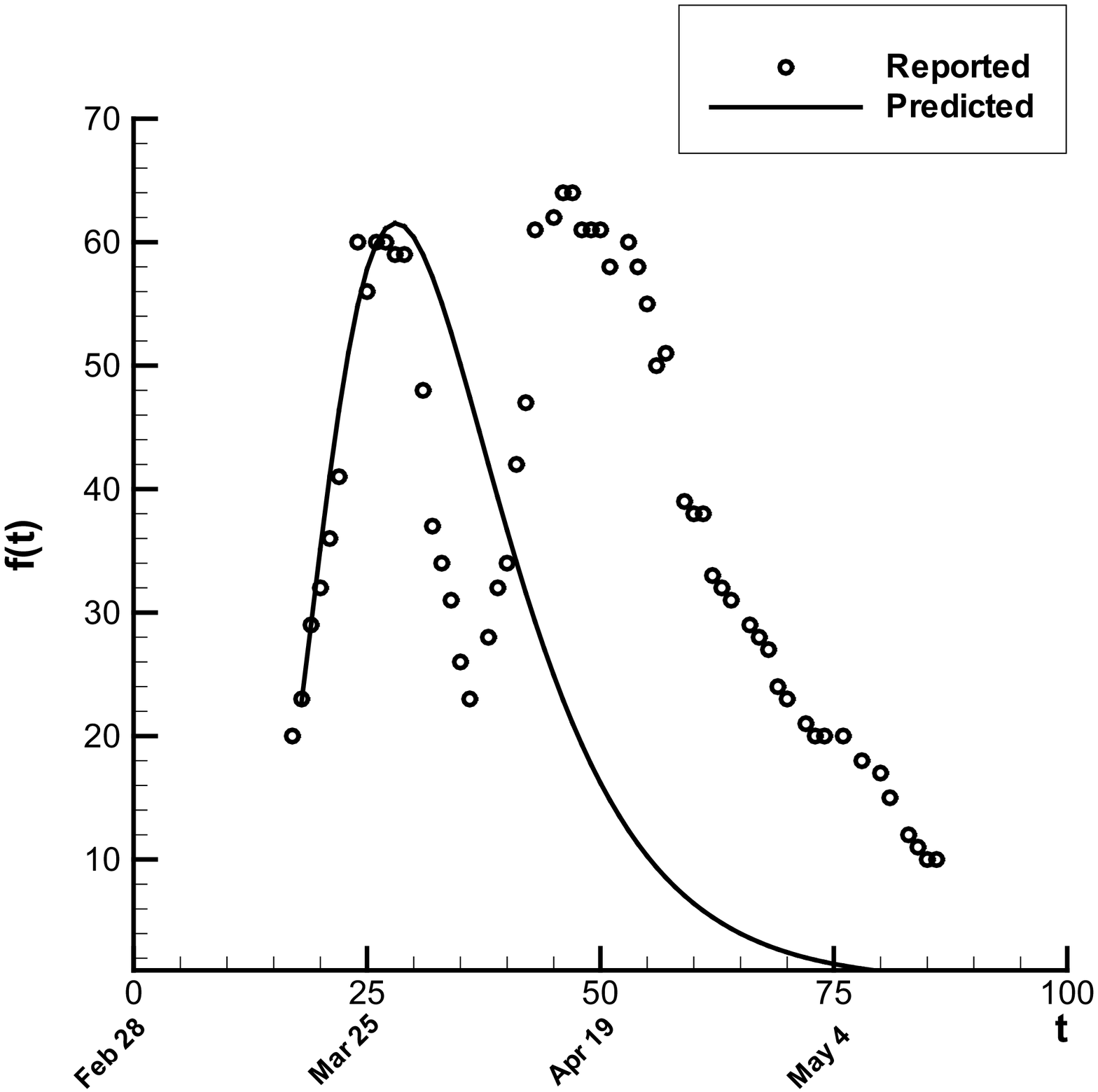}%
\caption{Comparion between predicted and reported cases for Singapore.}
\label{fg8}
\end{minipage}
\end{figure}

%

\begin{figure}
\begin{minipage}{0.45\textwidth}
\includegraphics[width=\textwidth]{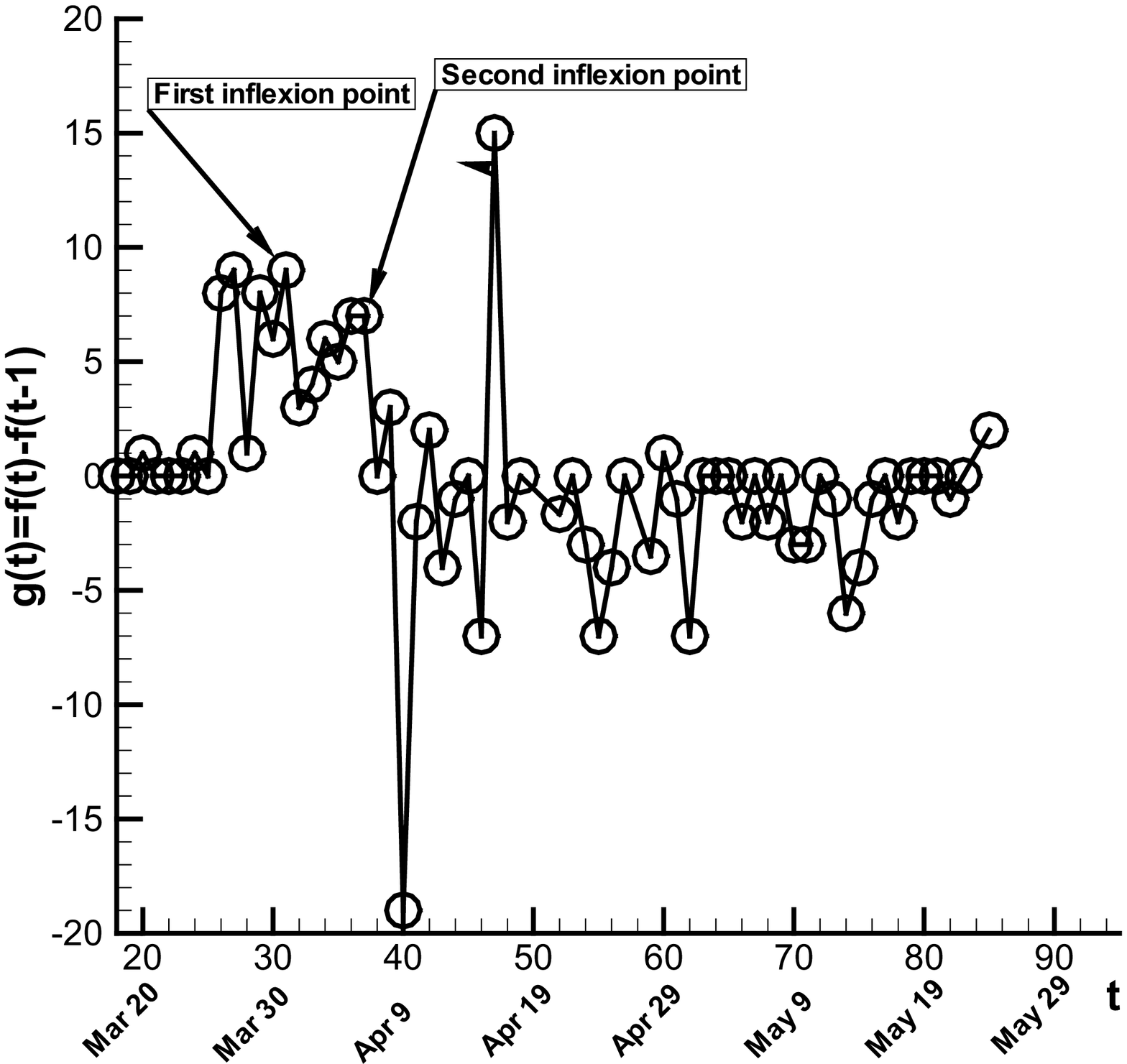}%
\caption{Inflexion points for Canada.}
\label{fg9}
\end{minipage}
\hfill
\begin{minipage}{0.45\textwidth}
\includegraphics[width=\textwidth]{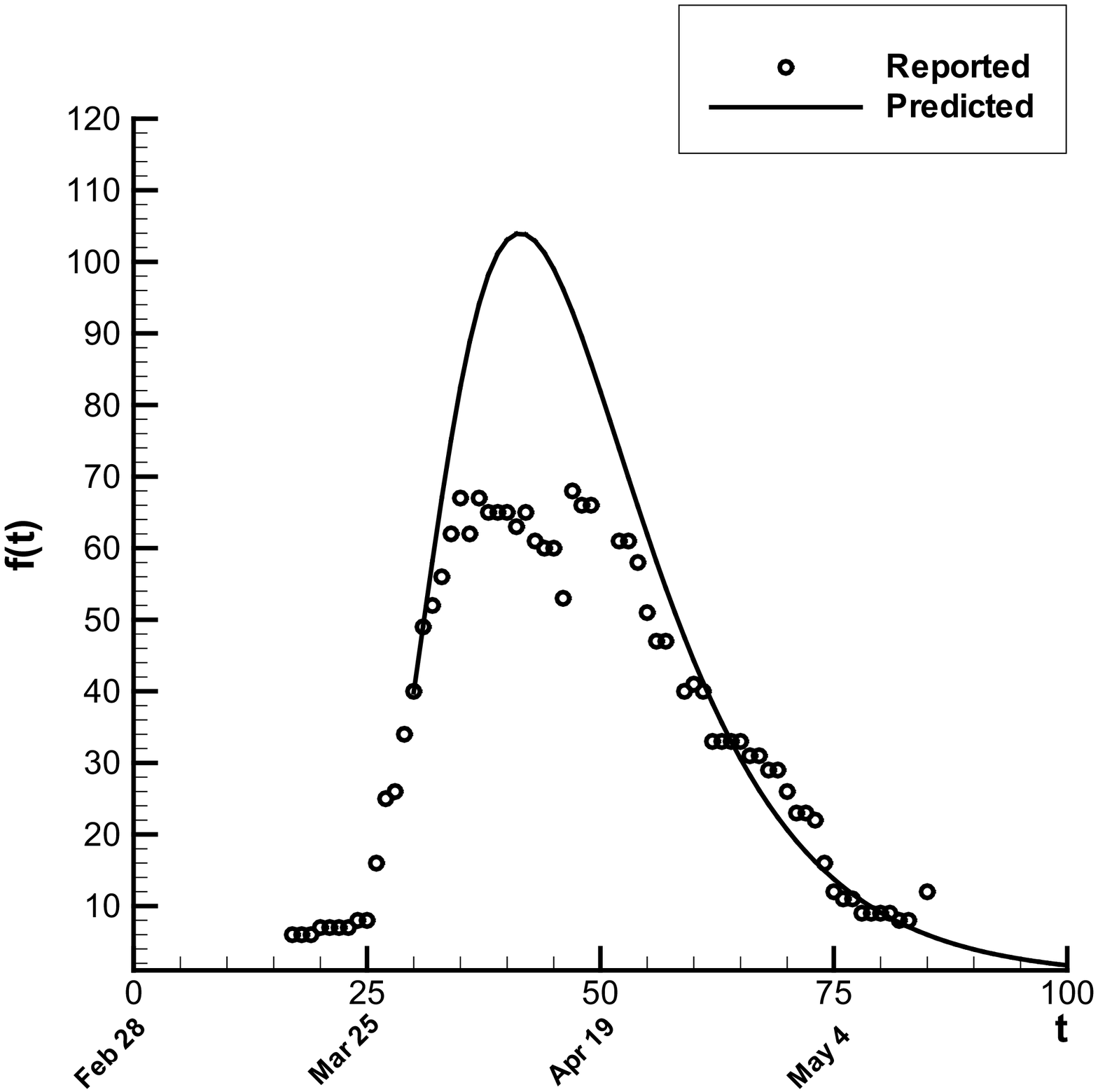}%
\caption{Comparison between predicted and reported cases for Canada.}
\label{fg10}
\end{minipage}
\end{figure}

For Singapore we observe three distinct inflexion points
and two maximums (Fig. \ref{fg7}). The prediction using the information at
the first inflexion point fits well to the most part of the first
peak (Fig. \ref{fg8}). For Canada we observe two inflexion points (Fig. \ref{fg9}) and when the information of the first inflexion point is used the
prediction reproduces well the lower part of the observed curve but fails to
predict the peak value (Fig. \ref{fg10}).

%

In summary, when the number is small, the error is large, showing that the
thermodynamic approach is more accurate when the system is larger.

\subsection{Comparison between the theoretical value of $\protect\sigma $
and the reported value of $\protect\sigma $}

It is interesting to note that the best fit value of $%
\sigma $ using the reported data is close to the theoretical one $(\sigma
=0.408)$ (Table 2). In fitting $\sigma $, the date $D$
(counting from the starting date) and the maximum value $H$ are fixed to be
the values given by the reported data (third and fourth columns) so that
only $\sigma $ is fitted. In the second column, the
starting date is approximately the date when the first
case was introduced into the region. The outbreak for the epidemic is
assumed to take place within at most ten days so the best fit $\sigma $ is obtained by using two epidemic starting days (date with
the introduction of the first case and latest possible
outbreak date). The range of best fit $\sigma $ (fifth column) is very close to the theoretical value $0.408$
for Beijing and is not significantly different from the
theoretical value for the other cities or regions.

\begin{table}
\begin{center}
\caption{Best fit value of $\sigma $ for several regions using the least square method, compared to the theoretical value $\sigma =0.408$. The agreement is reasonable. }
\label{tab2}
\begin{tabular}{|l|l|l|l|l|}
\hline
{\small Regions} & {\small Starting date} & {\small Date }$D$ & {\small %
Maximum }$H$ & Best f$^{{\small \cdot }}$it $\sigma $ \\ \hline
{\small Beijing} & {\small Mar 25/(10 days later)} & {\small May 15} &
{\small 1991} & 0.349/0.462 \\ \hline
{\small Hong Kong} & {\small Feb. 20/(10 days later)} & {\small Apr. 14} &
{\small 960} & 0.285/0.343 \\ \hline
{\small World} & {\small Feb. 20/(10 days later)} & {\small May 12} &
{\small 3700} & 0.273/0.32 \\ \hline
Mainland {\small China} & {\small Mar 25/(10 days later)} & {\small May 12}
& {\small 3068} & 0.307/0.404 \\ \hline
{\small Hebei} & {\small Apr. 17/(10 days later)} & {\small May 13} &
{\small 161} & 0.367/0.357 \\ \hline
Singapore & March 1{\small /(10 days later)} & Apr. 15 & 64 & 0.409/0.491 \\
\hline
Canada & Feb. 25{\small /(10 days later)} & April 8 & 84 & 0.297/0.368 \\
\hline
\end{tabular}
\end{center}
\end{table}

\begin{figure}
\centering
\includegraphics[width=0.6\textwidth]{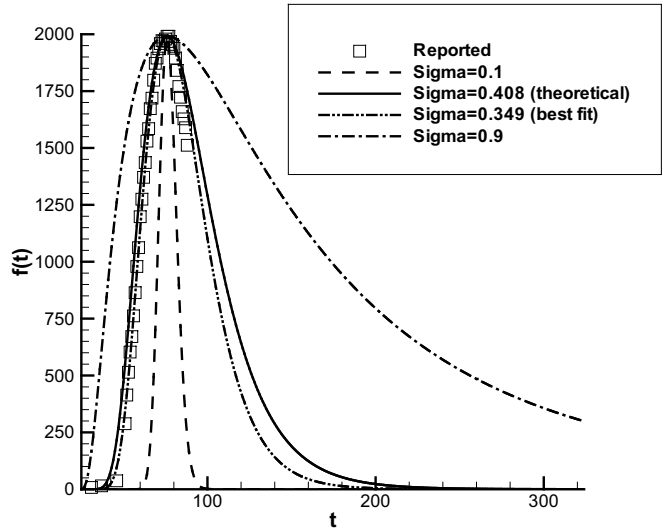}%
\caption{Role of $\protect\sigma$ on the correct reproduction of the curve $f=f(t)$.}
\label{fg11}
\end{figure}

One would wonder if the use of a value $\sigma $ far beyond the theoretical
value does not alter the curve significantly. In order to see that, we
display in Fig. \ref{fg11} the role of $\sigma $ on the correct
reproduction of the curve $f=f(t)$. The log-normal curves using the
thermodynamical value $\sigma =0.408$ and the best fit
value $\sigma =0.47$ are all close to the reported data. However, when $%
\sigma $ is significantly different from the
thermodynamical value, then the log-normal curve has a great departure from
the reported data, as can be seen from the curves using $\sigma =0.1$ and $%
\sigma =0.9$. This shows that the shape of the curve is quite sensitive to $%
\sigma $ and the theoretical value of $\sigma $ is indeed a rational one.

\section{Concluding remarks}
We have built a closed model for which we just need some data for the early
period to determine the inflexion point $L$, the number $%
f(L)$ and the increase rate $df(L)/dt$. The model is applied to predict the
number for $t>L$  and especially $D$ and $f(D)$ for the 2003 SARS and is hoped to work when the
system for SARS or similar epidemic spread involves multiscale interventions and constitutes a
thermodynamical system. Despite the possible uncertainty in the reported
data for $0<t<L$ and that the model does not require epidemic details such
as latent, incubation and infectious periods, the comparison\ between model
prediction and reported SARS data is still good enough for the cities or regions
where the epidemic is severe. The prediction for the case of Beijing is
remarkably well since the number of cases is very large. This shows that
when the system is large enough, the thermodynamic approach is more
accurate. The actual model has some difficulty to
exactly handle the case of multiple inflexion points.

The present model can be possibly used to predict
epidemics other than SARS once the communicable epidemics receive intensive interventions. The H7N9 avian influenza is actually of great concern \cite{ref-H7N9-1,ref-H7N9-2} and if unluckily this should spread rapidly, we expect the present model would be useful for predicting its spread.

For avian influenza, the observed severe symptom mainly includes high fever and pneumonia \cite{ref-H7N9-1,ref-H7N9-2}. It is interesting to note that, according to traditional Chinese meridian doctrine \cite{ref-jingluo,ref-jingluo2}, giving a pressure down or performing an acupuncture on specific acupuncture points on the meridian in a correct way by experts could be helpful to relieve or cure the corresponding symptoms (see Table \ref{tab3}).

\begin{table}
\begin{center}
\caption{The symptoms  and the corresponding meridian acupuncture points (reproduced from reference \cite{ref-jingluo2}). Pain is felt on the  meridian acupuncture points while a pressure is given on that points,  if the corresponding symptom exists}
\label{tab3}
\begin{tabular}{ccc}
\hline
  Symptom & Acupuncture (meridian) & Figure \\
  \hline
  high fever & zh$\bar{\mathrm{o}}$ngzh$\check{\mathrm{u}}$ (TE3; SJ3), & Fig \ref{fg_fever} \\
  & ti$\bar{\mathrm{a}}$nf$\check{\mathrm{u}}$ (L3; LU3),& \\
  & la$\acute{\mathrm{o}}$g$\bar{\mathrm{o}}$ng (P8; PC8),& \\
  & zh$\bar{\mathrm{o}}$ngch$\bar{\mathrm{o}}$ng (P9; PC9),& \\
  & d$\grave{\mathrm{a}}$zh$\grave{\mathrm{u}}$ (B11; BL11),& \\
  \hline
  pneumonia & d$\grave{\mathrm{a}}$nzh$\bar{\mathrm{o}}$ng (CV17; RN17), & Fig \ref{fg_lung} \\
  &  d$\grave{\mathrm{a}}$zh$\grave{\mathrm{u}}$ (B11; BL11), & \\
  & f$\grave{\mathrm{e}}$ish$\grave{\mathrm{u}}$ (B13; BL13), & \\
  & y$\acute{\mathrm{u}}$nm$\acute{\mathrm{e}}$n (L2; LU2), & \\
  & y$\acute{\mathrm{u}}$j$\grave{\mathrm{i}}$ (L10; LU10) & \\
  \hline
  \label{tab_pneumonia}
\end{tabular}
\end{center}
\end{table}

\begin{figure}
\centering
\includegraphics[width=0.7\textwidth]{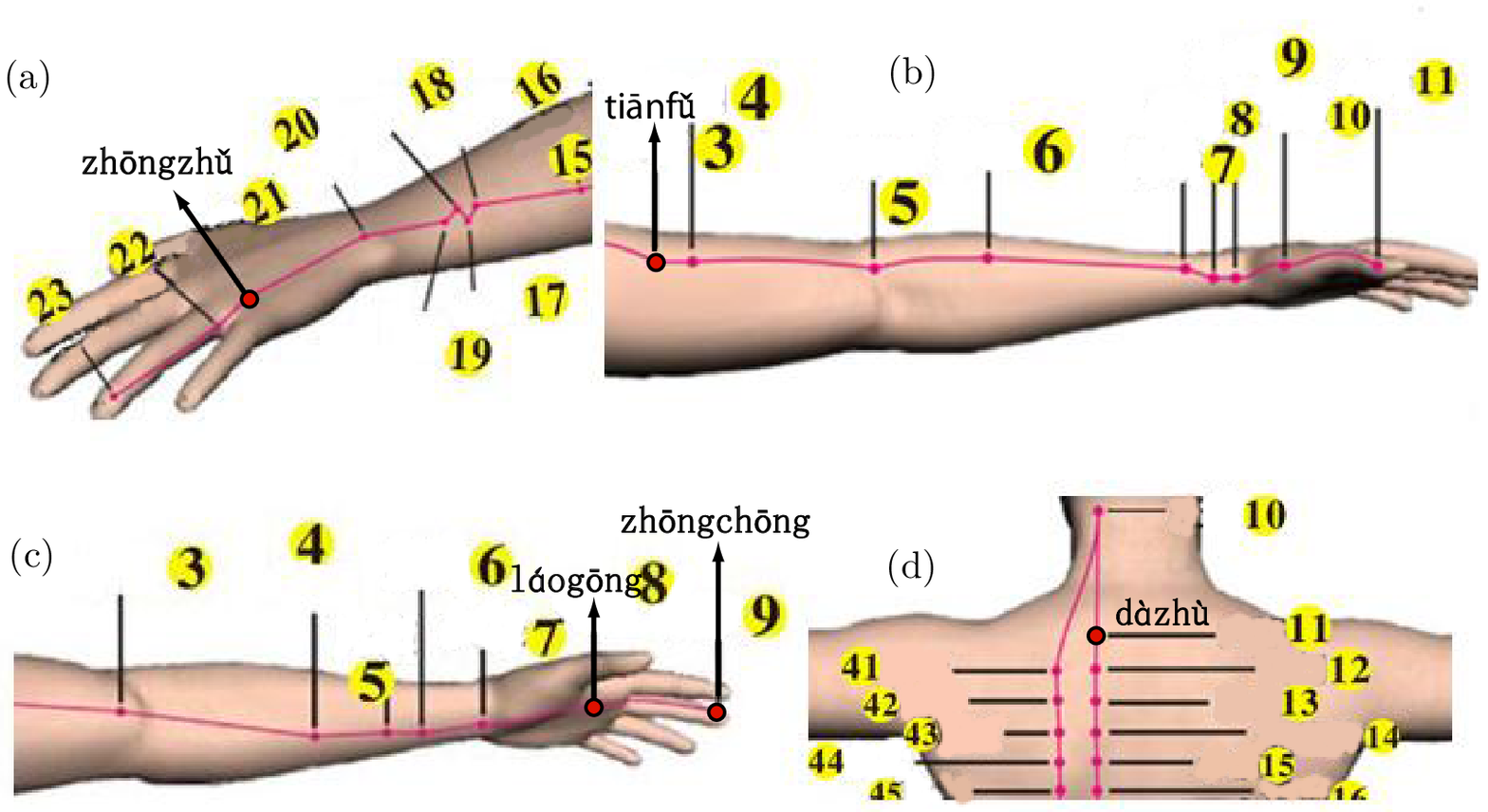}%
\caption{Acupuncture points related to high fever: (a) zh$\bar{\mathrm{o}}$ngzh$\check{\mathrm{u}}$ (shaoyang sanjiao meridian of hand), (b) ti$\bar{\mathrm{a}}$nf$\check{\mathrm{u}}$ (taiyin lung meridian of hand), (c) la$\acute{\mathrm{o}}$g$\bar{\mathrm{o}}$ng and zh$\bar{\mathrm{o}}$ngch$\bar{\mathrm{o}}$ng (jueyin pericardium meridian of hand), (d) d$\grave{\mathrm{a}}$zh$\grave{\mathrm{u}}$ (taiyang bladder meridian of foot).}
\label{fg_fever}
\end{figure}

\begin{figure}
\centering
\includegraphics[width=0.7\textwidth]{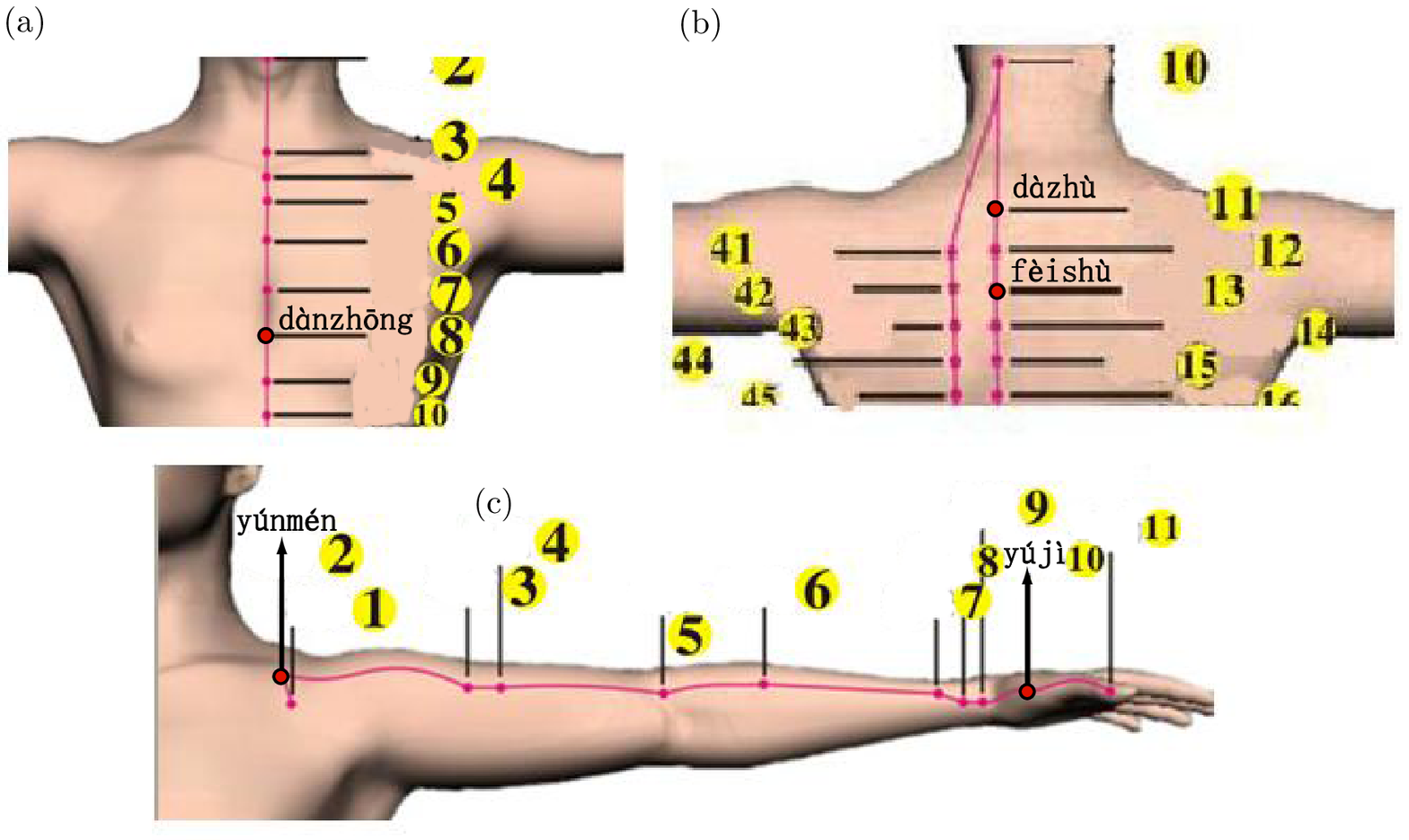}%
\caption{Acupuncture points related to pneumonia: (a) d$\grave{\mathrm{a}}$nzh$\bar{\mathrm{o}}$ng (ren meridian), (b) d$\grave{\mathrm{a}}$zh$\grave{\mathrm{u}}$ and f$\grave{\mathrm{e}}$ish$\grave{\mathrm{u}}$ (taiyang bladder meridian of foot), (c) y$\acute{\mathrm{u}}$nm$\acute{\mathrm{e}}$n and y$\acute{\mathrm{u}}$j$\grave{\mathrm{i}}$ (taiyin lung meridian of hand).}
\label{fg_lung}
\end{figure}

\section*{Acknowledge} This manuscript is updated from an unpublished manuscript originally written by Z.N.Wu during the SARS spreading in 2003.

\end{document}